\documentclass[]{article}

\title{A Note on Indexing Planar Point Sets for Approximate Bottleneck Distance Queries} 

\date{}
\author{Brendan Mumey}

\usepackage{amsmath}
\usepackage{amssymb}
\usepackage{amsthm}
\usepackage{url}

\newtheorem{lemma}{Lemma}
\newtheorem{definition}{Definition}

\newcommand{\db}{\mathbb{D}}
\newcommand{\tr}{\mathbb{T}}

\begin{document}

\maketitle

\begin{abstract}
The {\em bottleneck distance} is a natural measure of the distance
between two finite point sets of equal cardinality, defined as the minimum
over all bijections between the point sets of the maximum distance between any pair
of points put in correspondence by the bijection.
In this work, we consider the problem of building a data structure $\db$ that
indexes a collection of $m$ planar point sets (of varying sizes) and
supports nearest bottleneck distance queries: given a query point set $Q$ of 
size $n$, we would like to find the point set(s) $P \in \db$ of size $n$ 
that are closest in terms of bottleneck distance.
Without loss of generality, we assume that all point
sets belong to the unit box $[0,1]^2$ in the plane and focus on the
$L_\infty$ norm, although the techniques can also be used for other norms.
The main contribution is a {\em trie}-based data structure 
finds a $6$-approximate nearest neighbor in  $O(-\lg(d_B(\db,Q)) n)$ time,
where $d_B(\db,Q)$ is the  minimum bottleneck distance from $Q$
to any point set in $\db$.
\end{abstract}

\section{Introduction}

The bottleneck distance is a natural measure of the distance between 
two finite point sets of equal cardinality.  
The problem of computing the bottleneck distance
arises in geometric applications such
as comparing persistence diagrams in topological data analysis~\cite{Bubenik15}.
Bottleneck distance is defined between two point sets $P$ and $Q$ as
$$
d_B(P,Q) = \min_{h:P \rightarrow Q} \max_{p \in P} \lVert h(p) - p\rVert,
$$
where $h$ is a bijection and $\lVert \cdot \rVert$ is chosen 
as the $L_\infty$ norm, as this is common for the persistence diagram comparison application.
Given a database $\db$ of point sets, we can also define
$$
d_B(\db,Q) = \min_{P \in \db, |P|=|Q|} d_B(P,Q).
$$
Finally,
$$
\mbox{nearest}(\db, Q) = \{P : P \in \db, |P| = |Q|, d_B(P,Q) = d_B(\db,Q) \}.
$$
The problem considered in this work is to identify {\em approximate} nearest neighbor point sets $P$, whose bottleneck distance from $Q$ is within a constant factor of $d_B(\db,Q)$.
Without loss of generality, we assume that all point
sets belong to the unit box $[0,1]^2$ in the plane.  
We first describe a simple approach to represent point sets using strings.
This suggests using a {\em trie} data structure~\cite{DeLaBriandais:1959:FSU:1457838.1457895,Cormen} to store strings associated
with each point set in the database.  



\section{Related Work}

Bottleneck distance is closely related to the bipartite matching
problem, which can be solved by the classic maximum flow technique
of Hopcroft and Karp~\cite{HopcroftKarp}.  
The current best exact algorithm for bipartite matching of points in the plane is due
to Efrat {\em et al.}~\cite{Efrat2001} and
runs in $O(n^{1.5} \lg n)$ time for point sets of size $n$.

Earlier 
seminal 
work by Hefferman and Schira~\cite{Hefferman94} considered approximation
algorithms for 
the more general 
problem in which one of the point sets is mapped by
an isometry (translated, rotated, and possibly reflected) prior to being matched.
In the case of just computing the bottleneck distance, their methods
provide a $O(n^{1.5}(\epsilon/\gamma)^4)$ time algorithm to test if $d_B(P,Q) \le \epsilon$,
where the answer must be correct if 
$\epsilon \not \in [d_B(P,Q) -\gamma,d_B(P,Q) + \gamma]$.
A key idea in \cite{Hefferman94} is to 
check for bottleneck matchings using a maximum flow computation in graph
that arises from ``snap-rounding'' the point sets to their nearest point in grid.
Our approach uses a similar idea in which the maximum flow instance is
a planar graph (not true for \cite{Hefferman94}), so a recent improved
algorithm for multi-source, multi-sink maximum flow due to Borradaile~\cite{Borradaile11}
that runs in $O(n \log^3 n)$ time can be leveraged.

Bottleneck distance arises naturally in the comparison of persistence diagrams
in topological data analysis~\cite{Bubenik15}.
Fasy et al.~\cite{2018arXiv181211257F} 
consider the related problem of building a database
of persistence diagrams that permits approximate querying.
Their approach is 
also 
based on
representing point sets by snap-rounding
each point to neighboring grid points at each level in a multilevel grid data structure.  
All combinations of snap-roundings are considered and the 
resulting grid point configurations are stored in a database.  
Nearest point set queries are done in $O((n \log m + n^2)\log \tau )$ time, where
$\tau$ is number of grid levels used and $m$ is the number of point sets stored.
The resulting matches
are shown to provide a $6$-approximation to the nearest point set in the database.
Rather than using a hashing
scheme, it is possible to use a trie data structure (described in 
\S\ref{prelim}) to achieve $O(-\lg(d_B(\db,Q)) n)$ time queries.  

Approximation results are known for general bipartite matching in
metric spaces; in \cite{Agarwal14} the authors
show that for any $\delta > 0$, there is an algorithm that
computes a $O(\frac{1}{\delta^{\alpha}})$-approximate matching, 
where $\alpha = \log_3 2 \approx 0.631$
in $O(n^2 \log n \log^2\frac{1}{\delta})$ time.
A variation on minimum-distance
bottleneck matching, with the additional constraint that the matched edges cannot cross, was
recently shown to be NP-hard to approximate within a factor of less than $1.277$~\cite{Carlsson15}.

\section{Preliminaries}
\label{prelim}

Without loss of generality, all point sets are contained within
the unit box $B = [0,1]^2$ in the plane.  
Following the general approach of ~\cite{2018arXiv181211257F},
we recursively divide $B$ into finer grids.
The corner $(0,0)$ is designated as the {\em origin}.  
The four corners of $B$ are the grid points at level $1$.
The grid at level $d$ is subdivided by $2$ to form the grid at level $d+1$.  
Thus, 
level $2$ contains $9$ grid points and in general level $d$ contains $(2^{d-1}+1)^2$
grid points.  The grid length at level $d$ is $\delta_d = 2^{1-d}$.

Let $p$ be a point in or on $B$, for $d \ge 1$, 
we define $n_d(p)$
as the nearest level $d$ grid point to $p$, breaking ties by going in the S and/or W direction with respect
to $p$.  Observe that $n_d(p)$ is unique and that if $p$ is already a level $d$ grid point,
then $n_d(p)=p$.  We also define $n_0(p)$ as the origin.

Suppose $P$ is a point set to be stored in  $\db$.  For each $p \in P$, let
$$
n^4_d(p)=\{ g : \mbox{$g$ is a level $d$ grid point}, \lVert g - p \rVert_\infty< \delta_d \}.
$$
Note $|n^4_d(p)| \le 4$.  We consider all ways of snapping each $p$ to 
some grid point $\mbox{snap}(p) \in n^4_d(p)$.
\begin{definition}
A query point set $Q$ is
said to {\em hit} a point set $P$ at level $d$, if there is a snapping of $P$ such that
$| q : q \in Q, n_d(q) = g | = | p : p \in P, \mbox{snap}(p) = g|$ for all level $d$ grid points $g$.
\end{definition}
Versions of the following two lemmas appear in \cite{2018arXiv181211257F}, 
and a similar analysis is also found in \cite{Hefferman94}.
\begin{lemma}
\label{l1}
If $Q$ hits $P$ at depth $d$, then $d_B(P,Q) \le \frac{3}{2} \delta_d$.
\end{lemma}
\begin{proof}
Since $Q$ hits $P$, there is a snap-rounding of $P$ that produces
the grid configuration $n_d(Q) = \{ n_d(q) : q \in Q\}$ (repeats allowed); define a bijection $h: P \rightarrow Q$ by mapping each $p$ that
snapped to some grid point $g$ to a unique $q$ such that $g = n_d(q)$.
Then $\lVert h(p) - p \rVert_\infty = \lVert q - p \rVert_\infty \le \lVert q - n_d(q) \rVert_\infty + \lVert n_d(q) - p \rVert_\infty \le \frac{\delta_d}{2} + \delta_d$.  Thus $d_B(P,Q) \le \frac{3}{2} \delta_d$.
\end{proof}
\begin{lemma}
\label{l2}
If $Q$ does not hit $P$ at level $d$, then $d_B(P,Q) \ge \frac{\delta_d}{2}$.
\end{lemma}
\begin{proof}
Let $h: P \rightarrow Q$ be a bijection that
realizes $d_B(P,Q)$. We prove the contrapositive: Suppose $d_B(P,Q) < \frac{\delta_d}{2}$.  
Let $p \in P$ and $q = h(p)$.  Then 
$\lVert p - n_d(q) \rVert_\infty\le \lVert p - q \rVert_\infty+ \lVert q - n_d(q) \rVert_\infty 
\le  d_B(P,Q) + \frac{\delta_d}{2} < \delta_d$.
It follows that $n_d(q) \in n^4_d(p)$, and so snapping each $p$ to $n_d(q)$
provides a hit to $Q$ at level $d$.
\end{proof}
Suppose $d^*$ is the maximum depth at which there is a hit $P \in \db$ for a query point set $Q$.
Lemma~\ref{l1} implies $d_B(P,Q) \le \frac{3}{2} \delta_{d^*}$.
On the other hand, since no hits were
found at depth $d^*+1$, by Lemma~\ref{l2},
$d_B(\db,Q) \ge \frac{\delta_{d^*+1}}{2} = \frac{\delta_{d^*}}{4}$.
Thus $d_B(P,Q) \le 6 d_B(\db,Q)$, and so for a query point set $Q$,
the point set returned $P$ is
guaranteed to be a $6$-approximation to the nearest point set in $\db$.

\section{A Trie-based Data Structure}

We propose an indexing approach based on representing
configurations of grid points as strings.
We first define a string representation
for a single grid point at level $d$ as a length $d$ string and then
interleave $n$ such strings to represent a set of $n$ grid points in the level
$d$ grid.  The interleaving is done so that the string first describes the level $1$ 
grid points, then level $2$, etc.

Let $g$ be a grid point at level $d \ge 1$.  
We define $N_d(g)$ as the 
grid point neighbor at level $d$ directly north of $g$, provided this point belongs to the grid.
Define similarly for all eight principal compass wind directions and 
let $I_d(g) = g$ ($I$ for identity).  
We introduce a string encoding of any grid point $g$ at some level $d \ge 1$.
The string, $s_d(g)$ is
constructed in left-to-right order, in $O(1)$ time per symbol by ``walking'' in
the grid toward $g$, starting at the origin, following the grid points
$n_0(g), n_1(g), \ldots, n_d(g)=g$.  Observe that for $1 \le i \le d$,
\begin{equation}
n_i(g) = \mbox{dir}_i(g) (n_{i-1}(g)),
\label{eq:nw}
\end{equation}
where
$\mbox{dir}_i(g) \in \{I,N,S,E,W,NE,SE,NW,SW\}$.
Thus, we can compactly describe a level $d$ grid point $g$ 
by a unique string $s_d(g)$ of length $d$ over the nine symbols $\{\tt{I,N,S,E,W,NE,SE,NW,SW} \}$,
where the $i$th symbol indicates $\mbox{dir}_i(g)$.\\

We now consider how to use the above string encoding to represent
grid point configurations.  
Let $G$ be a set of $n$ grid points (repeats allowed) at level $d > 0$ and
let $S_d(G) = \{s_d(g) | g \in G \}$ be the set of length $d$ strings that encode
each grid point in $G$.  Consider $S_d(G)$ sorted into lexicographic
order, i.e. $S_d(G) = \{s_d(g_1) \le s_d(g_2) \le \ldots \le s_d(g_n)  \}$.
$S_d(G)$ can be encoded as a single interleaved string of length $nd$, defined as:
\begin{equation}
S_{d,G} = s_d(g_1)_1 \ldots s_d(g_n)_1\ \ s_d(g_1)_2 \ldots s_d(g_n)_2\ \  \ldots\ \  s_d(g_1)_n \ldots s_d(g_n)_n. 
\end{equation}
Notice that the first $n$ characters in $S_{d,G}$ describe the level $1$ nearest
neighbor grid points for $G$, the next $n$ characters describe the level $2$
nearest neighbor grid points for $G$ and so on.
Any distinguishable level $d$ grid point configuration $G$ is encoded uniquely by $S_{d,G}$.
The time required to generate $S_{d,G}$ is $O(dn)$ (e.g. by using 
radix sort).
\begin{lemma}
\label{nstring}
Let $p(G) = \{n_{d-1}(g) : g \in G\}$.
Then $S_{d,G} = [S_{d-1,p(G)}] s_d(g_1)_n \ldots s_d(g_n)_n$.  
\end{lemma}
\begin{proof}
This can 
be seen by noting that each string in $S_d(G)$ is formed from
a string in $S_{d-1}(G)$ with a single symbol appended to the end, so the lexicographic
sortings of $S_d(G)$ and $S_{d-1}(p(G))$ agree up to position $d-1$.
\end{proof}

A natural approach to storing a collection of point sets,
each represented as a string, is to use a {\em trie}-based 
data structure~\cite{DeLaBriandais:1959:FSU:1457838.1457895,Cormen}.

We first consider the database scheme proposed in \cite{2018arXiv181211257F},
in which many snap-roundings of each point set are stored and use
the aforementioned string representation and trie data structure.
To represent a point set $P$, snap-roundings
to grid point configurations (up to some maximum grid level $d_{\max}$) are
stored in a trie $\tr$. 
If $|P| = n$, there are $4^{d_{\max}n}$ such snap-roundings, although there
are potentially fewer distinguishable grid point configurations to store.  
Each snap-rounding at level $d$ represents a grid point 
configuration $G$ that must be
stored in $\db$; the string representation $S_{d,G}$ is used to represent each $G$.
\begin{lemma}
If $G$ is a snap-rounding configuration at level $d > 1$ for a point set $P$, then
then there is another snap-rounding configuration $G'$ of $P$ at level $d-1$ 
such that $S_{d-1,G'}$ is a prefix of $S_{d,G}$.
\end{lemma}
\begin{proof}
Let $G = \{g_1, \ldots g_n\}$ be a snap-rounding of $P$ at level $d > 1$.  
Each $g_i = \mbox{snap}(p) \in n^4_d(p)$ for some $p \in P$.
Let $g'_i = n_{d-1}(g_i)$.  Clearly, $g'_i \in n^4_{d-1}(p)$.  It follows
that the grid configuration $G' = n_{d-1}(G)$ will be snapped to
by $P$ in the level $d-1$ grid.  Furthermore, 
$S_{d,G} = [S_{d-1,G'}] s_d(g_1)_n \ldots s_d(g_n)_n$, by Lemma~\ref{nstring}.
\end{proof}

Each trie node will also store a pointer to 
the list of point sets (initialized to {\em null}).  
As snapped grid point configurations for $P$ are added to $\tr$, 
$P$ is appended to this list at each trie node that ``finishes'' describing a grid point
configuration for some level, e.g. if $|P|=k$, then a trie node at depth $dk$ 
describes a level $d$ grid point configuration for $P$.

The time required to add a new point set $P$ of size $n$ to $\tr$ is $O(4^{d_{\max}n} d_{\max} n)$,
since at most $O(4^{d_{\max}n})$ snapped grid configuration strings are stored and each 
is generated in $O(d_{\max} n)$ time.  The additional space requirement for $\tr$ is also
 $O(4^{d_{\max}n} d_{\max} n)$.

\subsection{Handling Queries}

Let $Q$ be a query point set of size $n$; our objective is to find those
$P \in \db$ that approximate \mbox{nearest}($\db$, Q), where the database $\db$
is represented using a trie $\tr$, as described above.
A query string $S_Q$ is constructed in left-to-right 
order in blocks of size $n$ as follows:
For each point $q \in Q$, we consider the sequence of grid points
$n_0(q), n_1(q), \ldots, n_{d_{\max}}(q)$; the sequence gets monotonically closer to $q$.
As before, we can represent this sequence as a string $s_{d_{\max}}(q)$, whose $i$th symbol is
$\mbox{dir}_i(q)$ and $S_{d_{\max}}(Q)$ is the collection of these strings for all $q \in Q$.
In order to produce the query string $S_Q$, $S_{d_{\max}}(Q)$ must be sorted lexicographically,
however this can be done {\em lazily} using radix sort.  First, $s_{d_{\max}}(q)_1$ is found
for all $q \in Q$ and the strings are sorted on index $1$.  The resulting sorted column
provides the first $n$ symbols in $S_Q$.  Next, the trie $\tr$ is searched on this block.  If there
is a hit, then the search continues to the next index position, the string symbols are computed at that position
(in $O(n)$ time) and the radix sort is continued at the next index.  This produces
the next size $n$ block of $S_Q$ and $\tr$ is probed from where the previous hit was found.
If $d^* \le d_{\max}$ is the maximum hit depth, then $d^*= -\lg(d_B(\db,Q))$ and
the query runs in $O(-\lg(d_B(\db,Q)) n)$ time.

\section{Discussion}

An approach to indexing planar point sets that supports
approximate nearest bottleneck distance queries using a {\em trie}-based
data structure to compactly represent
point configurations in a multi-level grid is described.
The obvious drawback 
is the exponential space
complexity; up to $4^{d_{\max}n}$ strings are stored for each point set of size $n$.
A natural question is whether a more space-efficient database scheme is possible.
It would also be interesting to consider if an indexing approach and
querying procedure can be found that permits one of the point sets
to be transformed by an isometry, such as done in \cite{Hefferman94}.





\bibliography{references}

\end{document}